# Magnetoelectric Exchange Spring for Spin-Wave Logic with Ultralow Energy Consumption

Aiji Wang[1,2]†, Mingming Li[3]†, Yuben Yang[1]†, Hanchen Wang[4]†, Yuelin Zhang[1,2]†, Liang Yang[1,2], Kai Liu[1,2], Sheng Cheng[5], Yang Yang[1,2], Yu Yun[1,2], Tianxiang Nan[6], Ka Shen[1,2], Tao Zhu[5,7], Xiufeng Han[7], Haiming Yu[4], Ce-Wen Nan[8], Jinxing Zhang[1,2]*

1. School of Physics and Astronomy, Beijing Normal University; Beijing, 100875, China.

2. Key Laboratory of Multiscale Spin Physics (Ministry of Education), Beijing Normal University; Beijing, 100875, China.

3. Key Laboratory of Materials Physics, Ministry of Education, School of Physics, Zhengzhou University; Zhengzhou, 450001, China.

4. Fert Beijing Institute, MIIT Key Laboratory of Spintronics, School of Integrated Circuit Science and Engineering, Beihang University; Beijing, 100191, China.

5. Spallation Neutron Source Science Center; Dongguan, 523803, China.

6.School of Integrated Circuits and Beijing National Research Center for Information Science and Technology (BNRist), Tsinghua University; Beijing, 100084, China.

7. Beijing National Laboratory for Condensed Matter Physics and Institute of Physics, Chinese Academy of Sciences; Beijing, 100190, China.

8. School of Materials Science and Engineering, Tsinghua University; Beijing, 100084, China.

† These authors contributed equally to this work

* Corresponding author. Email: jxzhang@bnu.edu.cn

**Abstract:**

Spin waves (magnons) are intrinsically free of Joule heating during their propagation in magnets. Exploring, understanding, and manipulating such quasiparticles not only helps reveal the exotic magnonic phenomena in condensed matter physics but also underpins the development of information devices with low energy consumption. However, achieving low-power and scalable control of spin-wave propagation remains a critical challenge. Herein, a magnetoelectric exchange spring has been artificially designed and constructed across an interface between ferroelectric $BaTiO_3$ and ferromagnetic $(La,Sr)MnO_3$ with low magnetic damping, in which polarization-modulated orbital reconstruction tailors local magnetic anisotropy and enables reversible control of spin-wave propagation. By successfully deviating from the Janovec-Kay-Dunn law, reversal polarization of engineered $BaTiO_3$ ($P_s$~1.2 μC $cm^{-2}$, $V_c$~0.4 V) is sufficient to switch the coherent magnon transport in the $(La,Sr)MnO_3$, indicating the magnetoelectric heterointerface may serve as a promising candidate for atto-joule control of spin-wave propagation (normalized to a 10×10 $nm^2$ device). Furthermore, taking advantage of amplitude, phase, and frequency to encode spin waves, all 16 compact Boolean logics with ultralow energy consumption have been designed. The artificial magnetoelectric exchange spring can be broadly applicable to various material choices, offering a versatile platform for developing energy-efficient and integrated magnonic devices.

Spin waves and their quasiparticle, magnons, are primary low-energy elementary excitations in magnetic systems, which serve as a bridge between microscopic spin interactions and macroscopic quantum phenomena [1-3]. Additionally, magnons obeying Bose-Einstein distribution enable spin transport that can avoid Boltzmann Tyranny, Joule heating and leakage disaster, which physically constrain the reduction of operating voltage and energy consumption in traditional semiconductor transistors [4-6]. The coherent magnons (spin waves) also can be used to encode and process information with low power consumption, high speed and compact structure via using their amplitude, phase, and frequency [7-11] [Fig. 1(a)]. Therefore, the ability to excite, propagate, and manipulate spin waves is a critical requirement for exploring hybrid magnonic quantum phenomena and developing energy-efficient information devices [7,12-16]. However, manipulation of spin wave with ultralow energy consumption at nanoscale has remained a formidable challenge [17].

Magnetoelectric effect provides the opportunity to control the spin degree of freedom or magnetic moment using electrical polarization that may be regulated with low energy consumption at nanoscale [18-25]. The total energy ($H_{total}$) of a magnetic moment consists of exchange interaction energy ($H_{ex}$), Dzyaloshinskii-Moriya interaction energy ($H_{DMI}$), dipole-dipole interaction energy ($H_{dip}$), anisotropy energy ($H_{an}$), Zeeman energy ($H_Z$), etc. ($H_{total}=H_{ex}+H_{DMI}+H_{dip}+H_{an}+H_Z$). The change in any of above energy terms via an efficient magnetoelectric effect, exhibiting as the variations in the magnitude and orientation of magnetization vector ($\boldsymbol{M}$), will modulate the amplitude ($A$), phase ($\varphi$) and frequency ($f$) of spin-wave propagation [26] ($\Delta\mathcal{F}(A, \varphi, f) = \Delta\mathrm{E}(\boldsymbol{M})$). Given that coherence enables the manipulation of an entire system via local modulation, it offers a novel paradigm to modulate the spin wave via a magnetoelectric heterointerface as seen in Fig. 1(b). However, atomic-scale designed heterointerfaces with such magnetoelectric coupling mechanisms in a low magnetic damping system at room temperature remain missing [6,19], inhibiting the realization of the ultralow-power control of spin-wave propagation and the design of energy-efficient information devices.

The previous results demonstrate that half-metal manganite $La_{0.67}Sr_{0.33}MnO_3$ (LSMO) exhibits low magnetic damping and enables the long-distance propagation of spin waves [27,28]. Using $BaTiO_3/La_{0.67}Sr_{0.33}MnO_3$ (BTO/LSMO) heterostructure as a model system, we report a non-volatile and complete on-off switch of spin-wave propagation by reversing the ferroelectric polarization at room temperature. A magnetoelectric exchange spring at the heterointerface identified by the layer-resolved polarized neutron reflectivity reconfigures the interfacial magnetic anisotropy, thereby effectively modulates the spin-wave propagation. Additionally, all 16 Boolean logics with ultralow operating energy and compact structures are designed based on the polarization-controlled spin-wave switch.

Pseudocubic (001)-oriented LSMO (~40 nm) thin film and (001)-oriented ferroelectric BTO have been epitaxially grown on $NdGaO_3$ (NGO) (110) substrates successively by pulsed laser deposition. The crystal structure has been studied using X-ray diffraction (XRD), confirming high-quality epitaxial growth of the heterostructure. High-resolution topographic image has been carried out by atomic force microscopy (AFM), which indicates that the heterostructure has a smooth surface with atomic steps and terraces. A clear box-in-box reversal of the out-of-plane ferroelectric domains of the heterostructure were observed by the piezoresponse force microscopy (PFM). The voltage-dependent PFM phase and amplitude signals confirm the reversible switching of polarization in the magnetoelectric heterostructure.

The bulk LSMO is trigonal crystal structure with a pseudocubic lattice constants $a_{pc}$ = 0.388 nm and the NGO is orthorhombic crystal structure with lattice constants $a$ = 0.543 nm, $b$ = 0.550 nm, and $c$ = 0.771 nm. The lattice mismatch between LSMO and NGO along the $[1\bar{1}0]$ and [001] directions of NGO are -0.47% and -0.70%, respectively [29]. The in-plane anisotropy in the magnetoelectric heterostructure creates a magnetic easy axis along the NGO $[1\bar{1}0]$ and magnetic hard axis along the NGO [001], which are identified by vibrating sample magnetometer (VSM) and reported before [29]. The ferromagnetic resonance (FMR) spectra of the magnetoelectric heterostructure have been measured. The low magnetic damping (~0.0019) in the metallic system indicates that the magnetoelectric heterostructure can serves as a great platform for polarization-mediated spin-wave propagation.

The propagating spin-wave spectroscopy (PSWS) has been investigated under the propagating direction (wave vector, **k**) perpendicular to the external magnetic field (**H**), so-called Damon-Eshbach (DE) geometry. Spin waves are excited and detected by two parallel nano-stripeline transducers (NSLs) in the magnetoelectric heterostructure with a distance of 2 μm as schematically shown in Fig. 2(a). The widths of the spin-wave waveguide and the NSL antennas are about 20 μm and 300 nm as seen in Fig. 2(b). By applying **k** along the easy axis of the LSMO (NGO $[1\bar{1}0]$) as described in the schematic of the measurement in Fig. 2(a), room-temperature PSWS have been performed before and after polarization switching [Fig. 2(c)]. A large frequency shift of the propagating spin waves at a biased magnetic field of ~18 mT is observed when the polarization has been reversed from downward to upward direction. The corresponding line plots can be seen in the Fig. 2(d), where the propagation can be completely switched 'On' and 'Off' by reversing the ferroelectric polarization under the operation of spin waves at 1 GHz. In addition, the spin-wave group velocity decreases from 1.92 km $s^{-1}$ to 1.50 km $s^{-1}$ after the polarization reversal, which indicates the change in spin-wave dispersion. Such a phenomenon is reversibly observed when polarization has been reversed back. However, negligible change in either frequency or group velocity is observed before and after polarization reversal when the PSWS in the DE geometry has been collected by applying **k** along the hard axis of the LSMO (NGO [001]). These results indicate that the amplitude ($A$), phase ($\varphi$) and frequency ($f$) of spin-wave propagation can be modulated by polarization switching.

To understand the mechanism of polarization-switchable PSWS, magnetization and magnetic switching process before and after the polarization reversal over the whole sample were measured by VSM. When the external magnetic field is along the in-plane hard axis (NGO [001]), the magnetic moment and hysteresis have significantly changed after the polarization reversal from downward to upward direction. However, there shows negligible change by applying external magnetic field along in-plane easy axis (NGO [$1\bar{1}0$]). Such behavior indicates a strong anisotropic magnetoelectric effect in the heterostructure. In order to further understand this, the reciprocal space mapping (RSM) measurements of NGO ($33\bar{2}$) and (240) planes have been carried out. When the polarization is reversed from downward to upward, the lattice constants of the heterostructure have ignorable change, which indicates that the strain effect is negligible. It is well recognized that the change of polarization direction can induce atomic displacements and polarization-determined orbital reconstruction across the heterointerface [30-33]. In upward-polarized BTO, the upward displacement of Ti ions suppresses interfacial orbital hybridization, intensifies the occupancy of Mn-$d_{x^2-y^2}$electronic orbital, and thus enhances the in-plane magnetic anisotropy of LSMO [32,33].

To clarify the magnetic structure, layered-resolved magnetism across the magnetoelectric heterointerface have been collected by the polarized neutron reflectivity (PNR). Figs. 3(a) and 3(b) show the results of PNR measurements at room temperature with BTO polarization upward and downward, respectively. The inset plots the spin asymmetry (SA), SA=($R^{++}$-$R^{--}$)/($R^{++}$+$R^{--}$), where the $R^{++}$ and $R^{--}$ are the reflectivity for spin parallel and antiparallel to the external field, respectively. Considering that there is only change in the BTO polarization, the profile of the nuclear scattering length density (NSLD) is the same during the process of fitting these two PNR data. Fig. 3(c) shows the nuclear and magnetic depth profiles of the sample under different BTO polarizations. By analyzing the PNR results, it is determined that the magnetization of LSMO is about 300 emu $cc^{-1}$, in consistency with the one obtained from VSM. A low magnetic scattering length density (MSLD) is observed with a thickness of ~13 nm, indicating that the moment vector rotates near the BTO/LSMO interface as shown in Fig. 3(c). The distribution of the moment vectors in LSMO layer with an upward polarization in BTO is shown in Fig. 3(d). The magnetic moments collected from the experimental results of PNR spiral from the easy to hard axis in the plane, shows an exchange-spring magnetic structure [34,35]. In contrast, there is no inhomogeneous distribution of magnetization observed in LSMO when BTO polarization is downward. Analogue to the magnetic exchange spring [34, 35], we define such a polarization-controlled spin structure as a magnetoelectric exchange spring. Based on the understanding of the magnetoelectric structure across the heterointerface, the magnetic moment rotation processes controlled by the polarization are shown in Fig. 3(e).

The aforementioned findings indicate that the magnetic anisotropy at the heterointerface controlled by the polarization, modulates magnetoelectric exchange spring, and thereby switches the spin-wave propagation. The theoretical calculation using Landau-Lifshitz-Gilbert equation also demonstrates that the spin-wave frequency is shifted when the magnetoelectric exchange spring forms around the critical magnetic field, which is consistent with our experiments.

Based on the above polarization-switched spin-wave propagation, logic devices can be designed where spin wave (voltage) serves as output (input) signal. The Boolean logics are the foundation of digital computing. Among them, the minimum computationally complete logic basis {Material implication (IMP), FALSE} can be considered as a bridge between simple Boolean logic and complex human reasoning, which was proposed by Russell and Whitehead in 1910 [36] and experimentally achieved by Hewlett-Packard (HP) Laboratories using two memristors in 2010 [37]. Thanks to the non-volatile control and the diverse information encoding of spin waves that offer greater flexibility for device design, the IMP Boolean logic can be implemented with a highly simple structure using one waveguide device as shown in Fig. 4(a). The LSMO bottom electrode and the Au top electrode fabricated between the NSL antennas are set as Inputs “A” and “B”, respectively. Defining the Input high (low) voltage level as logic 1 (0) and the Output high (low) spin-wave signal amplitude at a fixed frequency as logic 1 (0), the IMP Boolean logic can be realized by setting “A = 0, B = 1” as the initial state. The output results as well as polarization states under different inputs are presented in right panel of Fig. 4a, which corresponds to the truth table of the IMP operation.

Benefited by the magnetoelectric exchange spring across the heterointerface, an ultralow power consumption per operation of the logic can be achieved by engineering crystal symmetry and interfacial density of states, as well as scaling the thickness of BTO. The coercive voltage ($V_c$) and saturated polarization ($P_s$) decreases linearly with the decreasing thickness from 40 nm to 20 nm, which successfully deviates from the Janovec-Kay-Dunn scaling law [38]. Moreover, by introducing anisotropic biaxial compressive strain at the heterointerface, the saturated polarization of $BaTiO_3$ could be further reduced [39], thereby further lowering the switching energy barrier. The polarization-voltage hysteresis (*P*-*V*) loop in the magnetoelectric capacitor with 20 nm BTO layer shows that a $V_c$ of ~0.4 V and a $P_s$ of ~1.2 μC $cm^{-2}$ (Fig. 4b). Consequently, the IMP logic device have been experimentally fabricated and reversible operated using a low energy consumption. Fig. 4c further experimentally demonstrates the sequential operation of the IMP logic device under different inputs. Micromagnetic simulations demonstrate that spin-wave propagation can still be effectively tuned when the critical dimensions are scaled down to 10×10 $nm^2$, indicating the energy consumption of each operation may be reduced to 0.96 aJ ($2P_sV_c$ which refers to the benchmarking methodology from Intel) [20,40]. Under the same critical dimensions, the energy consumption per operation in such scalable device structure is lower than the one designed by other CMOS and Beyond-CMOS devices [37]. Moreover, the rest 15 Boolean logics with ultralow energy consumption are also designed in a single core structure by modulating the spin-wave amplitude and phase at a fixed frequency. Different logics can be reconfigured in the same structure by adjusting the bias of driving signals. We may define such spin-wave Boolean logics based on polarization-controlled magnetoelectric exchange spring as MES-Wave logic, which may provide considerable potentials for information processing via non-von Neumann architectures, such as Neuromorphic computing, in-memory computing, etc. Furthermore, the polarization-controlled magnetoelectric exchange spring provides a promising platform to switch the coherent magnons within topological spin textures that have been well studied in LSMO (e.g., skyrmions, bimerons) [27,41] and explore hybrid magnonic quantum phenomena (e.g., magnon-photon, magnon-phonon interactions) [42,43], thereby providing a brand-new condensed matter for future quantum technologies.

## Conclusions

In conclusion, a magnetoelectric exchange spring across the magnetoelectric heterointerface with low magnetic damping has been achieved so that the spin-wave propagation can be completely switched on/off with an ultralow operating energy. Benefitted by information encoded via using amplitude, phase and frequency, MES-Wave logic may unlock more feasible approaches for designing and interconnecting functional modules, which can be promising building blocks for meeting the unprecedented demands of the future information technologies (artificial intelligence, Internet of Things, etc.) with ultrahigh computational power and ultralow power consumption. Such a universal physical paradigm can be extended to a broader range of material choices, offering a versatile platform for exploring emergent phenomena arising from the interactions between magnons and other (quasi)particles and advancing their applications in information and quantum devices.

**Acknowledgments**

This work is supported by the National Natural Science Foundation of China (52225205, 12550004, 12304117, 12404119, 12304073, U23A20366) National Key R&D Program of China (2021YFA0718700), the Beijing Natural Science Foundation (240008), the Guangdong Provincial Quantum Science Strategic Initiative (Grant Nos. GDZX2501006).

**Author contributions**

† These authors contributed equally to this work

* Corresponding author. Email: jxzhang@bnu.edu.cn

A.W. and J.Z. conceived and designed the experiments. A.W. M.L. B.Y. and J.Z. are responsible for the sample preparation and PFM characterization. A.W. M.L. L.Y. K.L. T.N. and X.H. designed and fabricated the devices. A.W. M.L. H.W. and H.Y. conducted the VNA measurements and voltage-gating experiments and analyzed the data. B.Y. and Y.Y. conducted the structural characterization. A.W., B.Y. and M.L. performed VSM and FMR and analyzed the data. S.C. and T.Z. performed PNR and analyzed the data. K.S. provided critical theoretical discussion and calculation. C.N. and J.Z. provided useful insights all through the project. J.Z. supervised the experimental study. A.W., M.L., and J.Z. wrote the paper and discussed with all authors.

## Figures

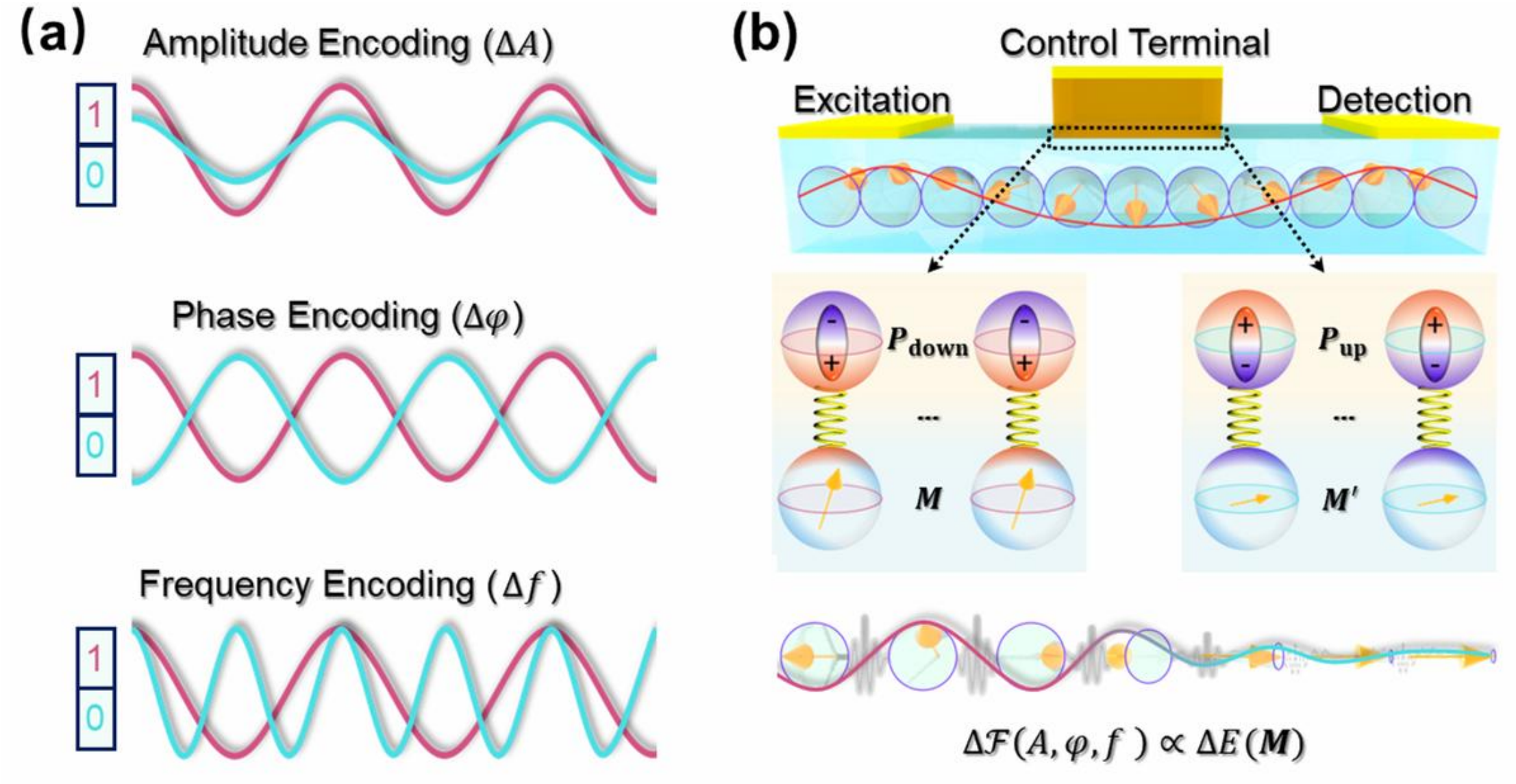

**FIG. 1. The physical paradigm of spin-wave device with ultralow operating energy across a magnetoelectric heterointerface.** (**a**) Illustration of various encoding scheme with spin waves as information carriers. (**b**) The physical paradigm for manipulating spin waves in magnetoelectric heterointerfaces. The interfacial magnetization vector ($\boldsymbol{M}$) is controlled by switching the ferroelectric polarization ($\boldsymbol{P}_{\mathbf{down}}$ and $\boldsymbol{P}_{\mathbf{up}}$). Since the coherent transport of spin waves, the change of interfacial magnetization vector ($\boldsymbol{M}$) can modulate the amplitude ($A$), phase ($\varphi$) and frequency ($f$) of the spin-wave propagation.

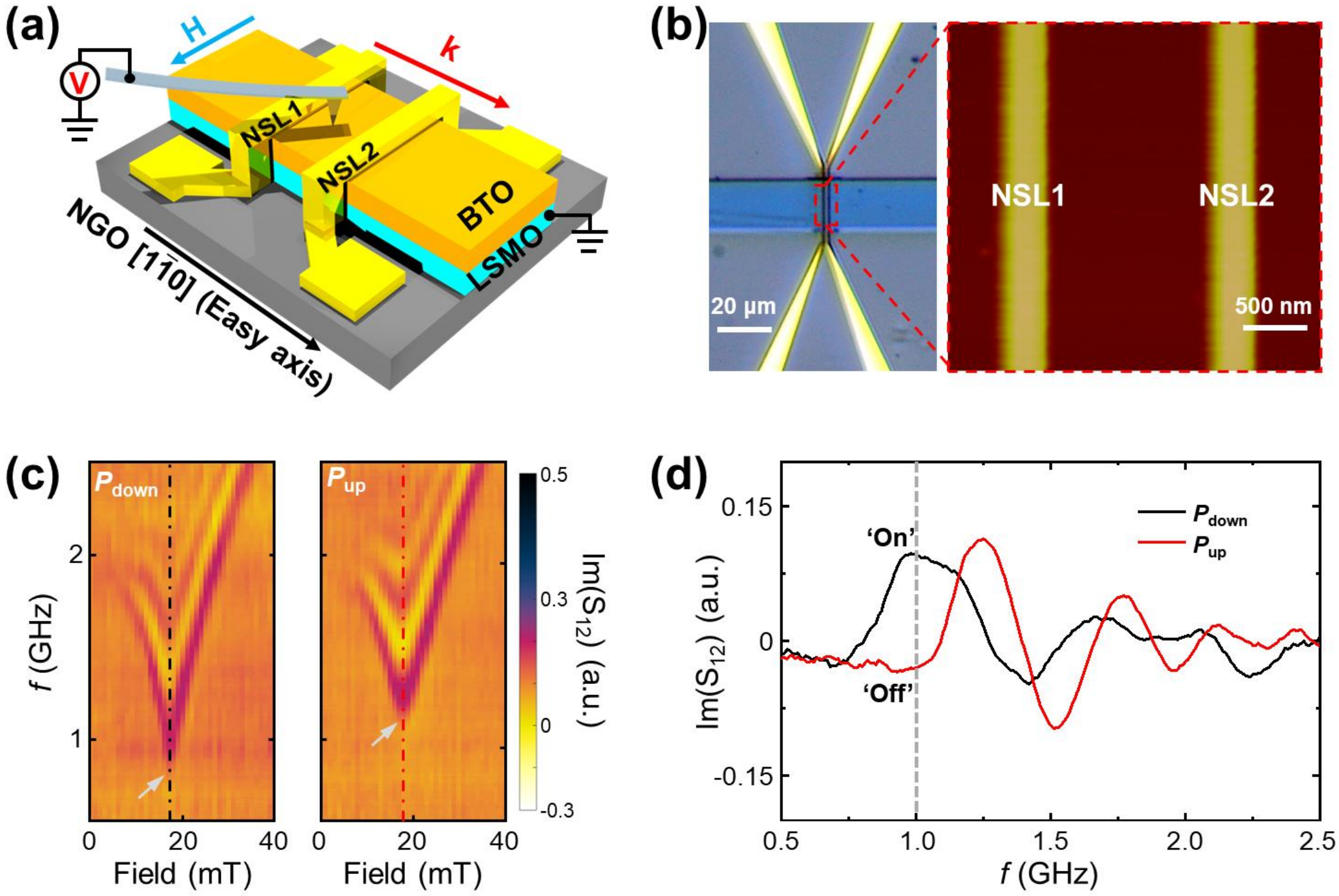

**FIG. 2. The propagating spin-wave spectroscopy (PSWS) switched by ferroelectric polarization.** (**a**) Schematic of PSWS measurement, spin waves excited/detected by the NSL antennas and propagate in the magnetoelectric heterostructure. The wave vector (**k**) is perpendicular to the external magnetic field (**H**). The PFM tip is used to control the direction of ferroelectric polarization. (**b**) The optical photograph and the AFM image of NSL antennas. (**c**) The PSWS of the heterostructure with polarization downward and upward. The grey arrows point out that the obvious frequency shift has been achieved after the polarization switches from downward to upward. (**d**) Line profiles of the PSWS, corresponding to the red and black lines in (c), where the propagation of spin waves at 1 GHz is completely switched 'On' and 'Off' by electric field.

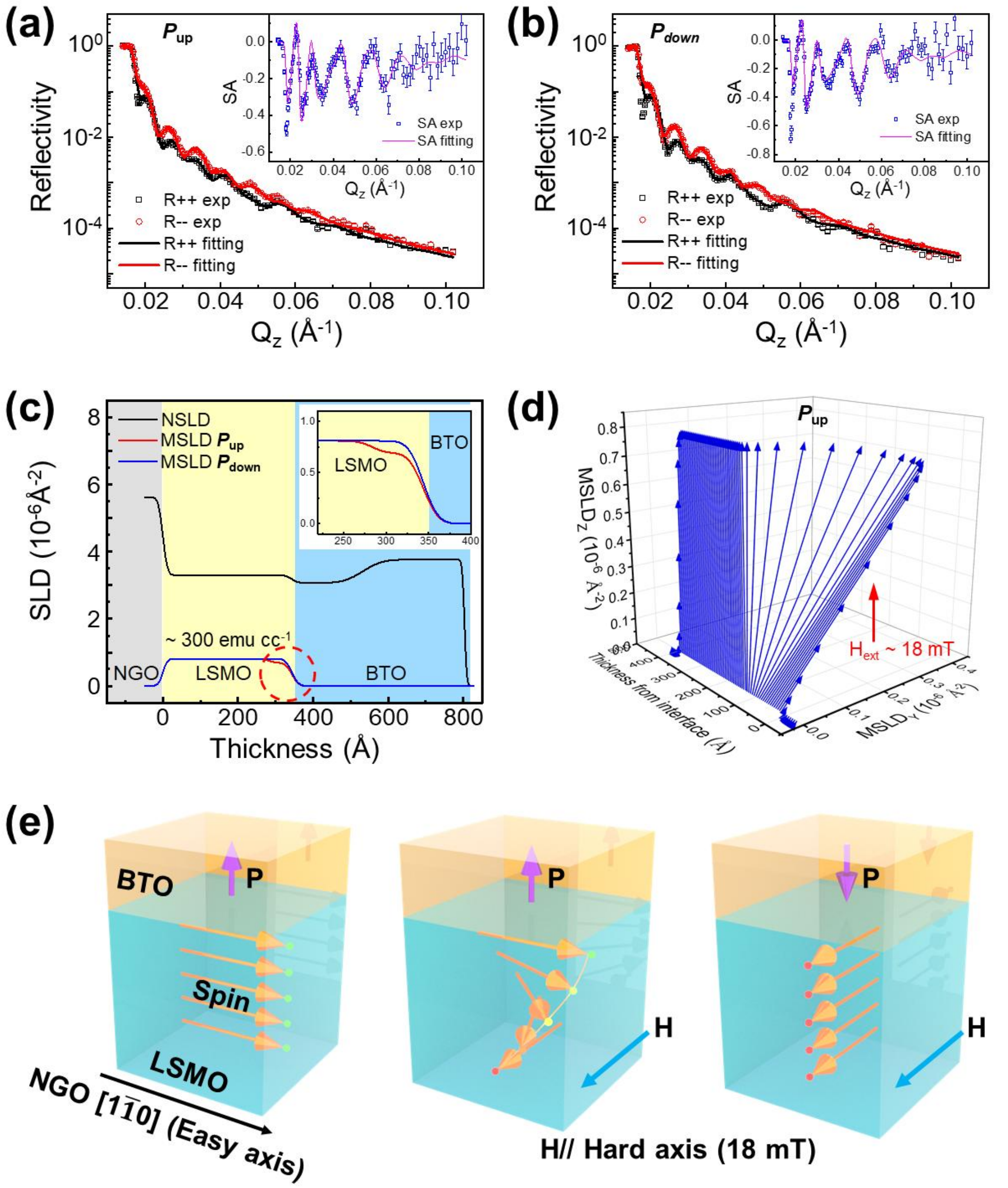

**FIG. 3. Polarization control of the magnetoelectric exchange spring.** (**a**), **(b)** Polarized neutron reflectivity (PNR) results of the magnetoelectric heterostructure with the polarization upward (a) and downward (b) at 18 mT magnetic field along the hard axis. The inset is the spin asymmetry. (**c**) The nuclear and magnetic depth profiles of the sample with different BTO polarizations. The magnetization of the LSMO near the interface has been regulated by the BTO polarization, which is zoomed-in and shown in the inset figure. (**d**) Magnetic moment vector distribution of the magnetoelectric heterointerface when the polarization is upward. The magnetic moment at the interface has pinned along the easy axis and the exchange spring magnetic structure forms. (**e**) The schematic illustration of the magnetoelectric exchange spring controlled by the polarization under a magnetic field applied along the in-plane hard axis.

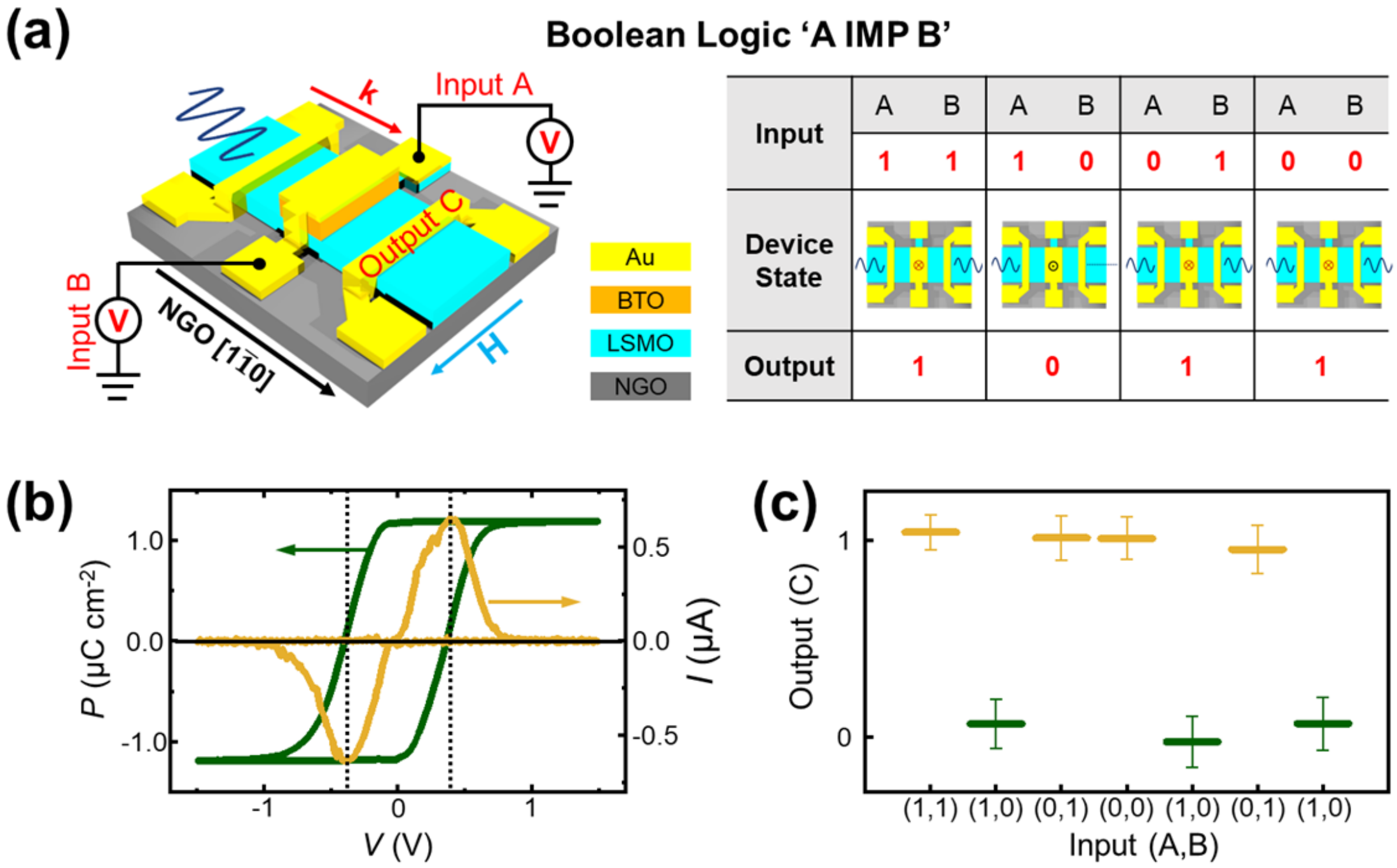

**FIG. 4. Magnetoelectric exchange spring spin-wave logic with ultralow operating energy. (a)** Schematic of the IMP logic device based on polarization-controllable spin waves, the bottom and top electrodes are set as Input "A" and "B" respectively, and the receiving antenna is set as Output "C". The truth table as well as polarization states of the IMP Boolean logic. (**b)** Polarization-voltage hysteresis loop and current-voltage curve of 20 nm BTO layer for estimating the operating energy. (**c)** The experimental results of the IMP under sequential operation.